\documentclass{aa}

\usepackage{graphicx}
\usepackage{txfonts}
\usepackage{lipsum}
\usepackage{subcaption}  
\usepackage{lscape}  
\usepackage{placeins} 
\usepackage{orcidlink}   
\usepackage{color}
\usepackage{natbib,twoopt}
\usepackage[hyphenbreaks]{breakurl}
\usepackage{hyperref}  
\bibpunct{(}{)}{;}{a}{}{,}            
\definecolor{cobalt}{rgb}{0.06, 0.2, 0.65}
\definecolor{rgreen}{RGB}{0, 128, 0}
\definecolor{sapred}{rgb}{0.5098039,0.1411765,0.2}

\hypersetup{
colorlinks=true,
linkcolor=teal,
citecolor=teal,
urlcolor=sapred
}
\makeatletter
  \newcommandtwoopt{\citeads}[3][][]{\href{http://adsabs.harvard.edu/abs/#3}
    {\def\hyper@linkstart##1##2{}
     \let\hyper@linkend\@empty\citealp[#1][#2]{#3}}}
  \newcommandtwoopt{\citepads}[3][][]{\href{http://adsabs.harvard.edu/abs/#3}
    {\def\hyper@linkstart##1##2{}
     \let\hyper@linkend\@empty\citep[#1][#2]{#3}}}
  \newcommandtwoopt{\citetads}[3][][]{\href{http://adsabs.harvard.edu/abs/#3}
    {\def\hyper@linkstart##1##2{}
     \let\hyper@linkend\@empty\citet[#1][#2]{#3}}}
  \newcommandtwoopt{\citeyearads}[3][][]
    {\href{http://adsabs.harvard.edu/abs/#3}
    {\def\hyper@linkstart##1##2{}
     \let\hyper@linkend\@empty\citeyear[#1][#2]{#3}}}
\makeatother

\usepackage{bm}
\usepackage{soul}
\usepackage{algorithm,algorithmic}
\makeatletter
\newcommand{\HEADER}[1]{\ALC@it\underline{\textsc{#1}}\begin{ALC@g}}
\newcommand{\ENDHEADER}{\end{ALC@g}}
\makeatother
\usepackage{amsmath,pdftexcmds}

\newcommand{\appropto}{\mathrel{\vcenter{
  \offinterlineskip\halign{\hfil$##$\cr
    \propto\cr\noalign{\kern0.5pt}\sim\cr\noalign{\kern-3.5pt}}}}}
    
\DeclareFontFamily{U}{cbgreek}{}
\DeclareFontShape{U}{cbgreek}{m}{n}{
        <-6>    grmn0500
        <6-7>   grmn0600
        <7-8>   grmn0700
        <8-9>   grmn0800
        <9-10>  grmn0900
        <10-12> grmn1000
        <12-17> grmn1200
        <17->   grmn1728
      }{}
\DeclareFontShape{U}{cbgreek}{bx}{n}{
        <-6>    grxn0500
        <6-7>   grxn0600
        <7-8>   grxn0700
        <8-9>   grxn0800
        <9-10>  grxn0900
        <10-12> grxn1000
        <12-17> grxn1200
        <17->   grxn1728
      }{}
\DeclareRobustCommand{\qoppa}{ \text{\usefont{U}{cbgreek}{\normalorbold}{n}\symbol{19}} }

\makeatletter
\newcommand{\normalorbold}{\ifnum\pdf@strcmp{\math@version}{bold}=\z@ bx\else m\fi}
\makeatother

\newcommand{\sco}{Sco~X-1}
\newcommand{\die}{PSR~J1023+0038}

\graphicspath{{./}{figures/}}
\usepackage[normalem]{ulem}

\begin{document}

\title{A semi-coherent search for optical pulsations from Scorpius X-1}

\titlerunning{A semi-coherent search for optical pulsations from Scorpius X-1}
\authorrunning{R. La Placa et al.}

   \author{Riccardo {La Placa}\inst{1}\fnmsep\thanks{\email{riccardo.laplaca@inaf.it}}
   \orcidlink{0000-0003-2810-2394}
        \and Alessandro Papitto\inst{1} \orcidlink{0000-0001-6289-7413}
        \and Giulia Illiano\inst{2,1} \orcidlink{0000-0003-4795-7072}
        \and Filippo Ambrosino\inst{1} \orcidlink{0000-0001-7915-996X}
        \and Christian Malacaria\inst{1}
        \and Luigi Stella\inst{1} \orcidlink{0000-0002-0018-1687}
        \and Paola Leaci\inst{3,4} \orcidlink{0000-0002-3997-5046}
        \and Pia Astone\inst{3}
        \and Cristiano Palomba\inst{3}
        \and Sara Motta\inst{2,5} \orcidlink{0000-0002-6154-5843}
        \and Adriano Ghedina\inst{6}
        \and Massimo Cecconi\inst{6}
        \and Francesco Leone\inst{7,8}
        \and Manuel González\inst{6}
        \and Héctor Pérez Ventura\inst{6}
        \and Marcos Hernandez Diaz\inst{6}
        \and José San Juan\inst{6}
        }

   \institute{INAF -- Osservatorio Astronomico di Roma, Via Frascati 33, I-00078, Monte Porzio Catone, Italy
            \and INAF -- Osservatorio Astronomico di Brera, Via Bianchi 46, I-23807, Merate (LC), Italy
            \and INFN, Sezione di Roma, I-00185 Roma, Italy
            \and Università di Roma ‘La Sapienza’, I-00185 Roma, Italy
            \and University of Oxford, Department of Physics, Astrophysics, Keble Road, OX1 3RH Oxford, United Kingdom
            \and Fundación Galileo Galilei -- INAF, Rambla J.A.Fernández P. 7, 38712, B.Baja (S.C.Tenerife), Spain
            \and Dipartimento di Fisica e Astronomia, Sezione Astrofisica, Università di Catania, Via S. Sofia 78, I-95123 Catania, Italy
            \and INAF -- Osservatorio Astrofisico di Catania, Via S. Sofia 78, I-95123, Catania, Italy\\
            }

   \date{Received July 30, 2025}
 
  \abstract
  {

  The emission of continuous gravitational waves (CWs) possibly explains why pulsars spinning with a period shorter than a millisecond have not been observed so far. Neutron stars accreting mass at the highest rates are the most promising targets for a search for CWs, because a strong emission of gravitational waves is required to balance the torque exerted by mass accretion onto the neutron star. Detecting coherent pulsations in the electromagnetic emission maximizes the search sensitivity, but has so far not been successful for most of the brightest accreting neutron stars. Here, we present the first search for pulsations in the optical band from the brightest accreting neutron star known, Sco X-1. To this end, we tailored semi-coherent search strategies to data obtained over four years, for a total of $\sim$$56$~ks, by the SiFAP2 fast photometer mounted at the Telescopio Nazionale Galileo (TNG). These searches are especially suited to analysing long observations of systems for which only limited knowledge on the orbital parameters is available, and involve joining coherent analyses on shorter segments without connecting the spin phase between them. The large count rates afforded by an optical telescope and the efficiency of the search strategy employed allowed us to set an upper limit of $9 \times 10^{-5}$ to the pulsed amplitude, which is lower by a factor of four with respect to previous searches in the X-ray band. We also show that the application of semi-coherent searches to SiFAP2 observations of the first detected optical millisecond pulsar, \die, could have preceded its detection in the radio band. These results highlight the role played by high-time-resolution optical observations in performing deep searches of quickly rotating pulsars.

  }


\maketitle
\nolinenumbers
\section{Introduction}

Accretion of mass and angular momentum transferred from a low-mass companion star in a binary system can spin a neutron star (NS) up to a rotational frequency of hundreds of hertz. Soon after the discovery of rotation-powered millisecond pulsars \citep[MSPs;][]{Backer1982}, low-mass X-ray binaries (LMXBs) were proposed as their progenitors \citep{Alpar1982,Radhakrishnan1982}. Accreting X-ray millisecond pulsars in transient LMXBs \citep[][]{patruno2021,disalvo2022}
demonstrated {that disc accretion can actually spin up a NS to such periods} \citep[][]{Bisnovatyi-Kogan1976,wijnands1998}. Observations of transitions between accretion- and rotation-powered emission on timescales of less than a few days eventually proved the evolutionary link between these two classes of NS \citep{archibald2009,papitto2013,Bassa2014,papitto2022}.

{Keplerian break-up limits the maximum spin of a NS} at $\sim$$1$~kHz or above, depending on the equation of state describing the rotating compact object \citep{Cook1994, Lattimer2007}. In principle, detecting a very fast, sub-millisecond spinning NS would immediately rule out some of the proposed models \citep[see e.g.][and references therein]{Bhattacharyya2016}. However, the observed distribution of spin frequencies of 
MSPs sharply cuts off at $\sim$700~Hz \citep{Chakrabarty2003, Chakrabarty2008, Hessels2008,Papitto2014,Patruno2017}.
The strongly frequency-dependent spin-down torque related to the emission of continuous gravitational waves (CWs) might explain the lack of very quickly spinning NSs \citep{Bildsten1998,Andersson2005}: elastic and/or magnetic deformations of the distribution of the accreted matter onto the NS surface, instabilities from bar-mode and r-mode oscillations, and free precession are some of the mechanisms proposed to drive off a NS from axis-symmetry and emit CWs (see \citealt{Sieniawska2019} and references therein). 
Besides selection effects, alternative models to explain the lack of sub-millisecond pulsars include magneto-dipole rotation and/or propeller spin down occurring when the secular mass-accretion phase ends altogether \citep{Burderi2001,Tauris2012} or when it decreases during the quiescent phases of X-ray transients \citep{Stella1994,Bhattacharyya2017}, and standard magnetic disc accretion torque theory for given values of the pulsar magnetic field and accretion rate \citep{Andersson2005}.

The possibility that CW emission might limit the spin rate of MSPs prompted several search attempts with the latest generations of GW interferometers, with no detection reported to date \citep[see e.g.][]{LVK2022c,LVK2022d,LVK2022a,LVK2022b,LVK2022e,Whelan2023,Steltner2023,LVK2025}. Assuming that CW emission balances the spin-up imparted to the NS by mass accretion, the brightest accreting NSs in LMXBs are widely considered the best candidates for CW detection (see \citealt{Watts2008} and references therein).
With an accretion rate close to matching the Eddington rate at a distance of a few kiloparsecs \citep[][]{Bradshaw1999,Bradshaw2003,McNamara2005}, Scorpius X-1 stands out in this regard. 
In addition, some of its orbital parameters were measured with good accuracy through optical observations \citep[][]{Wang2018,Killestein2023}, reducing the volume of the parameter space that a coherent signal search has to investigate (see Sect.~\ref{sec:intromethods}). However, deep searches for a coherent spin modulation at radio \citep{Taylor1972} and X-ray energies \citep{Wood_1991,Ubertini1992,vaughan94,Galaudage2022} were not successful. This significantly hampered the sensitivity of CW searches \citep{LVK2007,LVK2019,LVK2022e,Zhang2021,Whelan2023,Amicucci2025}, as accurate knowledge of the position in the sky, the binary parameters, and the spin frequency and its evolution is required to maximize their sensitivity.

Recently, the visible band emerged as a promising alternative channel to search for fast coherent signals from MSPs. SiFAP2 \citep{Ghedina2018}, a fast optical photometer operated from the 3.6~m Telescopio Nazionale Galileo (TNG), detected optical pulsations from a transitional (\citealt{Ambrosino2017,Papitto2019,Illiano2023}; see also \citealt{Zampieri2019,Karpov2019}),
 an accreting \citep{Ambrosino2021}, and a rotation-powered MSP \citep{Papitto2025}. 
The minimum detectable pulsed amplitude, $A$, depends on the number of photons in the observation, $N_{\gamma}$, through $A \propto N_{\gamma}^{-1/2}$ \citep[e.g.][]{Wood_1991,vaughan94}.
An instrument such as SiFAP2 from the TNG records around a $\text{few}\times10^5$ photons per second from a source with an apparent magnitude of $V\sim12$--$13$ mag, such as \sco\ (see Table~\ref{tab:SCOobs}). As a result, searches for coherent signals in the optical band are slightly more sensitive for a given exposure time than, for example, X-ray telescopes, which in turn typically observe approximately $\text{a few}\times10^4$ photons per second, at most \citep[see e.g.][]{LaMonaca2024}, although the relative contribution from the NS expected in the X-ray band is higher than in the optical one. For the two optical MSPs that also show X-ray pulsations, SAX J1808.4-3658 and \die, the pulsed fractional amplitudes in the X-ray band are approximately ten times higher than in the optical band: however, taking observations made with SiFAP2 and the PN instrument on XMM-Newton as an example, the ratio between their photon count rates in the optical and X-ray bands are $\sim$150 and $\sim$4000, outweighing the offset in pulsed amplitude \citep{Ambrosino2017,Ambrosino2021}.

Here, we present the first search for optical pulsations from {\sco} using SiFAP2/TNG observations. To maximize the sensitivity, we adapted semi-coherent search techniques (see Sect.~\ref{sec:intromethods}) to the case of optical data. Section~\ref{sec:method} describes our algorithm, which loosely follows the one applied in \citet[][]{Messenger2015}, whereas in Sect.~\ref{sec:sign} and~\ref{sec:upper_limits} we discuss the statistical properties of the combined detection statistics and the significance of both signals and upper limits. Section~\ref{sec:1023} shows a first application of the algorithm to data from a known pulsar, \die, and Sect.~\ref{sec:SCOX1} contains the analysis of 13 observations of \sco\ through the same method. Finally, in Sect.~\ref{sec:discussion} we review our results and discuss future applications.

\section{Coherent periodicity searches}
\label{sec:intromethods}
Searching for a fast coherent signal from LMXBs requires dedicated techniques to account for the drifts in the observed signal frequency caused by the orbital motion. As a result, a periodicity search with, for example, a Fourier power density spectrum (PDS), is either limited to a `simple search' on short time intervals, over which the binary frequency drift is negligible compared to the frequency resolution, or requires the preliminary correction of the photons' detection times for the periodic variation of the travel time at different phases during the orbit. While the sensitivity of the former approach is sub-optimal due to the reduced number of photons in the time series and the coarser frequency resolution, the latter can become computationally prohibitive if the orbital parameters are not known in advance with a good level of precision. The cost of a `blind search' by trial and error on a fine grid of possible combinations of orbital parameters 
scales with the length of the time interval coherently analysed as $\appropto T_{\mathrm{coh}}^7$.\footnote{We followed the convention of the astronomical community to refer to these as blind searches. In the gravitational-wave community, the same expression refers to all-sky searches that do not assume any prior knowledge about the source, not even its position, while `directed searches' would be used in our case \citep[see e.g.][]{Leaci2017, Wette2023}.} In addition, the minimum detectable amplitude scales with the number of trials as $A^2\propto \log{N_{tr}^2}$ \citep[e.g.][]{Wood_1991,vaughan94}.
`Acceleration' and `jerk searches' approximate the frequency drifts caused by the orbital motion over intervals short enough with respect to the binary period with a polynomial \citep[][]{Hertz1990,Johnston_1991,Wood_1991,Bagchi2013,Andersen2018}. As a result, they maintain the signal coherence over longer segments than simple searches, attaining greater sensitivity. However, these searches still analyse each segment separately.
Sideband searches represent an alternative approach to studying observations that are longer than the orbital period, complementarily to acceleration searches \citep[][]{Ransom2003}. They exploit the fact that orbital motion produces sidebands in the PDS around the spin frequency of the pulsar with a spacing equal to $T_{\mathrm{obs}}/P_{\mathrm{orb}}$, which can be then used to correct for the orbital modulation.

To make the best possible use of large datasets which span a time interval much longer than the binary period and obviate the enormous computational costs associated with the search of weak CW signals from systems with large parameter uncertainties, \citet[][]{Messenger2011} proposed to use `semi-coherent searches'.
This strategy was developed to search for CWs from isolated NSs \citep[][]{Brady2000} and has since been widely applied in searches focused on binary systems
\citep[][]{Pletsch2009,Astone2014,Leaci2015,LVK2017,Dergachev2019,Wette2023}. 
It involves a coherent analysis of separate segments of a long observation and the incoherent sum of their detection statistics, keeping track of the signal orbital evolution. Here, by `incoherent' we mean that information on the signal phase is lost between segments. Although less sensitive than fully coherent blind searches performed on the same observation, these techniques are computationally much more efficient, so that at fixed computational cost they permit us to search larger portions of the parameter space or reach higher sensitivities \citep[e.g.][]{Prix2012}. 
{Similar to the case of CWs, searching for binary gamma-ray pulsars relies on year-long data integration and naturally implemented semi-coherent strategies, yielding some detections in recent years \citep[][]{Pletsch2012Science,Nieder2020detection,Clark2021}.}
The method from \citet{Messenger2011} was also applied four times to X-ray data of LMXBs \citep[][]{Messenger2015,vandenEijnden2018,Patruno2018,Galaudage2022}.

In the next section, we describe in full the semi-coherent algorithm we adopted to analyse high-photon-count-rate datasets observed in the optical band.

\section{Method} \label{sec:method}

In the case of a NS in a binary system, any periodic signal at the spin frequency, $f$, is observed as modulated by the Doppler shift due to the projected velocity of the pulsar. Assuming a circular orbit, the detection time $t$ of a photon emitted at a given time $\tau$ is, after correcting for our position relative to the Solar System barycentre,
\begin{equation} \label{eq:tarr}
    t = \tau + \frac{a_1\sin(i)}{c} \ \sin \left[ \frac{2 \pi}{P_{\mathrm{orb}}} (\tau -T_{\mathrm{asc}}) \right] + \frac{d}{c} \ ,
\end{equation}
where $a_1\sin(i)/c = a_\perp$ is the projected semi-major axis of the NS orbit around the binary's centre of mass, $P_{\mathrm{orb}}$ is the orbital period, $T_{\mathrm{asc}}$ is the time at which the NS crosses the ascending node during its orbit, and $d$ is the binary's distance from the Solar System.
The observed frequency of the signal is
\begin{equation} \label{eq:niobs}
    \nu_{\mathrm{obs}}(t) = f \left(1 - a_\perp \Omega \cos{\left[ \Omega(t-T_{\mathrm{asc}}) \right]} \right) \ ,
\end{equation}
where we introduced $\Omega = 2\pi/P_{\mathrm{orb}} $ to simplify the notation. 
Therefore, in the PDS the signal can be spread across all the frequency bins that go from the minimum to the maximum of $\nu_{\mathrm{obs}}(t)$ across an orbit.

To take into account this frequency drift while keeping the problem computationally tractable, \citet{Messenger2011} proposed to divide the process into two steps. First, a coherent pulsation search is performed on $M$ separate segments of length $T_{\mathrm{seg}}$ out of a $T_\mathrm{span}$-long observation (the coherent step). Then, each orbital parameter combination of interest is explored by summing the powers found in different segments during the coherent step, at the frequencies produced by the signal modulation according to Eq.~(\ref{eq:niobs}) (the semi-coherent step). 

\subsection{The coherent step}

In the following, we always consider segments that are short compared to the orbital period of the source.
In each segment $m$, we can approximate the observed frequency of the signal by Taylor-expanding Eq.~(\ref{eq:niobs}) to the {$s^*$-th} order \citep[][after correcting one sign]{Messenger2011,Messenger2015}:
\begin{equation} \label{eq:nitaylor}
    \nu_{\mathrm{app}}^{(m)}(t) \simeq \nu_{1}^{(m)} + \sum_{s=2}^{s^*} \frac{\nu_{s}^{(m)}}{(s-1)!} \left(t- t_{\mathrm{mid}}^{(m)} \right)^s \, ,
\end{equation}
where $t_{\mathrm{mid}}^{(m)}$ is the mid-point of the segment, 
\begin{equation} \label{eq:nis}
   \nu_{s}^{(m)} = \left\{   \begin{array}{cr}   
   f \left(1 - a_\perp \Omega \cos{\left[\gamma^{(m)}\right]} \right)  \qquad    &\text{for } s=1    \\   
   -f a_\perp \Omega^s \sin{\left[\gamma^{(m)} + s\frac{\pi}{2} \right]} \qquad  &\text{for } s>1   
   \end{array} \right. 
\end{equation}
and $\gamma^{(m)} = \Omega(t_{\mathrm{mid}}^{(m)} - T_{\mathrm{asc}})$.
Similarly, the time correction from each PTA in the segment, $t_{j}^{(m)}$, to its related emission time at the NS can be approximated through
\begin{equation} \label{eq:tautaylor}
    \tau_{j}^{(m)} \simeq t_{\mathrm{mid}}^{(m)} + \frac{1}{f} \sum_{s=1}^{s^*} \frac{\nu_{s}^{(m)}}{s!} \left(t_{j}^{(m)} - t_{\mathrm{mid}}^{(m)} \right)^s \, .
\end{equation}

The uncertainty range in $\Omega$ and $a_{\perp}$ sets the range spanned by the various $\nu_{s}^{(m)}$ (see Eq.~\ref{eq:nis}), i.e. the boundaries of each dimension of the coherent space.
Then, we populated our coherent space by placing a grid of combinations of $\nu_{s}^{(m)}$ for each segment, whose spacing depends on the maximum fractional loss in power, $\mu^*$, that we allowed. We can associate with any given $\mu^*$ a coherent space metric, $g_{ij}$, which describes the distance between points in the space of vectors $\vec\nu = (\nu_1,\nu_2,\dots,\nu_{s^*})$ \citep[][]{Messenger2011}. 
The correct general form of the metric is found in Sect.~IV~C of \citet{Leaci2015}, but we are particularly interested in the diagonal terms, $g_{jj}$, which are a good approximation of its general form for short segments \citep[see][]{Messenger2011}, and the first four terms are\footnote{\citet{Leaci2015} expressed their short-segment coherent metric, $\tilde{g}^{SS,v}$, in terms of their rescaled dimensionless $v$ coordinates, while their $u$ coordinates are what we call $\nu_s$, i.e. our coherent search variables; the coherent metric we need, $g(\vec{\nu})$, is thus given by (with their notation) $\tilde{g}^{SS,u} \simeq \tilde{g}^{SS,v} \frac{\partial v_k}{\partial u_k} \frac{\partial v_{k'}}{\partial u_{k'}}$.}
\begin{equation} \label{eq:gdiag}
    g_{jj} = \frac{\pi^2 T_{\mathrm{seg}}^2}{3} \times\, \left(1, \ \frac{T_{\mathrm{seg}}^2}{60}  , \ \frac{T_{\mathrm{seg}}^4}{1344} , \ \frac{T_{\mathrm{seg}}^6}{172800} , \ \dots   \right) \, .
\end{equation}

We placed our points on hypercubic cells whose sides are then given by \citep[][]{Messenger2015}
\begin{equation} \label{eq:deltanis}
    \delta \nu_s = 2\sqrt{\frac{\mu^*}{s^* g_{ss}}}  \, ,
\end{equation}
and in the following we always use $\mu^* = 5\%$.
Therefore, we define $s^*$ as the largest order for which the total range found through Eq.~(\ref{eq:nis}) is larger than its related spacing following Eq.~(\ref{eq:deltanis}) would be.

On each grid point of the coherent space, we corrected the photons' detection times through Eq.~(\ref{eq:tautaylor}) and create $M$ matrices of coherent detection statistics, i.e. the power $\Lambda^{(m)}$, by calculating the Leahy-normalized PDS from the $\tau_{j}^{(m)}$ \citep[][]{leahy83}. All of the $\Lambda^{(m)}$ are saved for use in the semi-coherent step.

\subsection{The semi-coherent step} \label{sec:semicoherentstep}

Searches based on uncertain orbital parameters entail exploring a grid of possible orbital parameter combinations, which for brevity we call templates, within those uncertainties. 
The traditional method for the construction of such a grid requires that every point in the parameter space be `covered' at least by one template, although this becomes computationally heavy as the number of dimensions, i.e. the number of parameters, increases. As was the case for the coherent space, this implies that the grid spacing must be dense enough that any point is no further than a given distance from the closest template, which is a function of the maximum acceptable loss in signal power.

We tiled the orbital parameter space following \citet{Caliandro2012}, which provided exact expressions for the expected fraction of signal power recovered, $\varepsilon^2$, for a given distance in each dimension, thus creating a lattice. We allowed for a maximum loss for each dimension equal to $1 - \varepsilon^2 = 2.5\%$, so that, in the worst possible case where all three binary parameter,s $a_\perp$, $P_{\mathrm{orb}}$, and $T_{\mathrm{asc}}$, are as far as possible from the closest lattice point, we still retain $\varepsilon^6\sim92.7\%$ of the original power. Making the same choice, if we only search over two of the binary parameters, as is the case for the analyses discussed in Sects.~\ref{sec:1023} and \ref{sec:SCOX1}, we retain $\sim$$95\%$ of the power.

The search for higher-frequency coherent signals requires finer grids in all of the physical parameters with respect to low-frequency ones; therefore, the analysis is split in 100-Hz wide frequency bands. 
For each of them, we used the minimum spacings required ($\delta P_{\mathrm{orb}}, \,\delta T_{\mathrm{asc}}, \, \delta a_\perp $) across all segments at the maximum frequency of that band. 
The spacing in intrinsic spin frequency is taken to be the independent discrete Fourier one, $1/T_{\mathrm{seg}}$. 

For each template in the parameter space, i.e. each combination $\vec{\theta} = (f, P_{\mathrm{orb}}, a_\perp, T_{\mathrm{asc}})$, we can use Eqs.~(\ref{eq:nitaylor}) and (\ref{eq:nis}) to search for the corresponding nearest neighbour in the coherent space in each segment, hence defining a `track' in the coherent space. The sum of all the related powers, $\Lambda^{(m)}$, defines the combined statistics, $\Sigma$, whose properties are discussed in Sects.~\ref{sec:sign} and~\ref{sec:upper_limits}.
The full algorithm is summarized in Appendix~\ref{app:algo}.

\section{Significance of detections} \label{sec:sign}

The total detection statistic, $\Sigma$, is the sum of the powers found in the PDS of each of the segments. In the absence of a signal, and assuming pure Poissonian noise, $\Sigma$ is the sum of $M$ individual $\chi^2$ distributions with two degrees of freedom each, and therefore follows a $\chi^2$ distribution with $2M$ degrees of freedom.
The probability, $\mathcal{P}$, of obtaining a value of the summed statistic equal to or greater than $\bar{\Sigma}$ in a single trial in the case of noise alone is given by
\begin{equation}\label{eq:psingle}
    \mathcal{P}_{st}(\chi_{2M}^2\geq\bar{\Sigma}) = 1 - \frac{\gamma\left(M,\frac{\bar{\Sigma}}{2} \right)}{\Gamma(M)} = 1 - P\left(M,\frac{\bar{\Sigma}}{2} \right)  ,
\end{equation}
where $\gamma$, $\Gamma$, and $P$ are the lower incomplete, complete, and regularized lower incomplete gamma functions, respectively. Here, by single trial we mean a given sum of $M$ Leahy-normalized power spectra. 
The probability of obtaining at least one value of the detection statistic in the same range in the case of $n$ trials is thus
\begin{equation} \label{eq:pmulti}
    \mathcal{P}_n(\chi_{2M}^2\geq\bar{\Sigma}) = 1-(1-\mathcal{P}_{st}(\chi_{2M}^2\geq\bar{\Sigma}))^n = 1 - \left(P\left(M,\frac{\bar{\Sigma}}{2} \right)\right)^n \, ,
\end{equation}
and, consequently, once we choose a false-alarm probability, $\mathcal{P}_{\mathrm{FA}}$, that we are willing to accept as adequately small, the equivalent threshold in total detection statistic is given by
\begin{equation} \label{eq:sigmapfa}
    \Sigma^{\mathcal{P}_\mathrm{FA}} = 2P^{-1}\left(M,\left( 1 - \mathcal{P}_\mathrm{FA}\right)^{1/n}  \right) \, ,
\end{equation}
where by $P^{-1}$ we denote the inverse of the regularized lower incomplete gamma function, $P$, in Eq.~(\ref{eq:psingle}).
Strictly speaking, Eq.~(\ref{eq:pmulti}) only holds in the case of $n$ independent values of $\Sigma$ across our parameter space; since our template grid is dense in order to avoid missing possible signals, the noise realizations in templates that are close to one another could not be completely statistically independent. 
This can skew the probability distribution away from Eq.~(\ref{eq:pmulti}), undermining the statistical significance of signals above the threshold (see the example of \citealt{Nieder2019} for the $H$ statistic). However, as shown in Appendix~\ref{app:multitrial}, for the $\chi^2$ noise distributions analysed in this paper this leads to the thresholds obtained through Eqs.~(\ref{eq:pmulti})~and~(\ref{eq:sigmapfa}) actually being at most more conservative than necessary. We therefore used them without further corrections.

In principle, different combinations of binary parameters can produce the same evolution of the observed signal in the coherent space, i.e. the same track, owing to the finitely spaced grid in the coherent space. As a result, the number of unique tracks walked during our semi-coherent step, $N_{walked}$, can be smaller than the total number of combinations of spin and orbital parameters, $N_{comb}$. However, only in a handful of cases was $N_{walked}$ lower than $N_{comb}$, and the ratio $N_{walked}/N_{comb}$ was always higher than $99.9\%$ for the analyses presented in Sects.~\ref{sec:1023} and \ref{sec:SCOX1}. We thus considered $N_{comb}$ as a conservative upper limit on the number of statistically independent trials made in our search. However, we note that if the maximum acceptable losses in power in the two steps of the algorithm, $\mu^*$ and $1 - \varepsilon$, were set to very different values, the ratio $N_{walked}/N_{comb}$ would drop significantly: taking $N_{comb}$ in those cases could lead to a great overestimation of the number of independent trials carried out.
Since at each frequency interval we consider the spacing in both physical and coherent templates banks change, 
the final number of total trials is the sum of the trials in each frequency interval.

\section{Upper limit determination} \label{sec:upper_limits}
Should no summed power $\Sigma$ be found above the detection threshold, we can place an upper limit to the signal power, $\Sigma_\mathrm{signal}$, that might have been present in the data at any chosen confidence level. A detailed study of the probability distributions in power spectra was carried out by \citet{groth75}, and we used the procedures and renormalization outlined in \citet{vaughan94}, which we briefly summarize in the following, to derive upper limits both on signal power and fractional amplitude from our maximum detected power, $\Sigma_\mathrm{max}$. 

The joint probability density of measuring $\Sigma_\mathrm{meas}$ through the sum of $M$ power spectra, containing a signal of power $\Sigma_\mathrm{signal}$ is \citep[][renormalized]{groth75} 
\begin{equation}
\label{eq:grothpdf}
\begin{split}
 p_M(\Sigma_\mathrm{meas}, \Sigma_\mathrm{signal}) & = \frac{1}{2}\left(\frac{\Sigma_\mathrm{meas}}{\Sigma_\mathrm{signal}}\right)^{\frac{M-1}{2}} \exp\left[-\frac{\Sigma_\mathrm{meas}+\Sigma_\mathrm{signal}}{2}\right] \times \\ 
 & \times I_{M-1}\left(\sqrt{\Sigma_\mathrm{meas}\Sigma_\mathrm{signal}}\right) \, ,
\end{split}
\end{equation}
where $I_{l}$ indicates the modified Bessel function of the first kind of order $l$ \citep[see e.g.][]{NIST_Handbook}. This is simply the non-central $\chi^2$ distribution first obtained by \citet{Fisher1928}, $\chi'^2$, with $2M$ degrees of freedom and non-centrality parameter equal to $\Sigma_\mathrm{signal}$; therefore, Eq.~(\ref{eq:grothpdf}) can be integrated providing the probability of obtaining a power up to $\Sigma_\mathrm{max}$ if a signal of power $\Sigma_\mathrm{signal}$ is present in the data \citep[][]{Continuous_univariate_distributions_vol2}: 
\begin{equation}
\label{eq:noncentralcdf}
\begin{split}
 \mathcal{P}(\chi_{2M}'^2(\Sigma_\mathrm{signal}) & \leq \Sigma_\mathrm{max}) = e^{-\Sigma_\mathrm{signal}/2}  \times \\ 
 & \times \sum^\infty_{j=0}  \frac{\left(\frac{\Sigma_\mathrm{signal}}{2}\right)^j}{j!} \mathcal{P}_{st}(\chi_{2M+2j}^2\leq\Sigma_\mathrm{max})  \, .
\end{split}
\end{equation}
Such process can be numerically inverted to obtain, for any choice of confidence level, $C$, the corresponding upper limit, $\Sigma_\mathrm{UL}$, on signal power within our observation \citep{vaughan94}: that is, a signal such that the probability of it producing a measured power less than or equal to $\Sigma_\mathrm{max}$ is $\mathcal{P}_\mathrm{FD} = 1 - C = \mathcal{P}(\chi_{2M}'^2(\Sigma_\mathrm{UL}) \leq \Sigma_\mathrm{max})$.

Finally, we can express the background-subtracted upper limit on the pulse amplitude as \citep[see e.g.][]{vaughan94,Bilous2019} 
\begin{equation} \label{eq:Aul}
    A_{\mathrm{ul}} = \frac{N_\mathrm{\gamma,tot}}{N_\mathrm{\gamma,tot}-N_\mathrm{\gamma,bkg}} \,  \frac{\sqrt{\Sigma_{\mathrm{UL}}/{N_\mathrm{\gamma,tot}}}}{\mathrm{sinc}\left(\frac{\pi}{2} \, \frac{\nu_{\mathrm{ul}}}{\nu_{\mathrm{Nyq}}}\right)} \, \frac{1.61}{\sqrt{(1-\mu^*) \,(\varepsilon^2)^r}}\, , 
\end{equation}
where $N_\mathrm{\gamma,tot}$ and $N_\mathrm{\gamma,bkg}$ are the total number of photons and the ones from the background alone, respectively, from all of the summed segments, while the factor 1.61 arises from the discrete Fourier spacing, and $r$ is the number of orbital parameters over which the search is performed.

\section{Test on PSR J1023+0038} \label{sec:1023}

To test the algorithm, we used it to reproduce the detection of optical pulsations from SiFAP2 data of a known pulsar in a binary system, \die. It was the first transitional MSP to be discovered \citep[][]{archibald2009}, and the first one for which pulsations were detected in the optical band \citep[][]{Ambrosino2017}. 

We chose an optical observation from the night starting on 30 January 2020 (UTC), for a total length of almost 19~ks and $427\,675\,147$ collected photons (target plus background). All photon detection times were corrected to the Solar System barycentre with the DE405 planetary ephemeris assuming $\alpha = 10^h23^m47.687198^s$ and $\delta = +00^\circ38\arcmin40.84551\arcsec$ as source coordinates \citep{Jaodand2016}. This was done through a custom pipeline which takes into account both the geometrical R{\o}mer delay and the relativistic Einstein and Shapiro delays from the Sun, neglecting effects due to atmosphere, parallax, and other planets, and was already successfully used in previous analyses of SiFAP2 observations \citep[e.g.,][]{Ambrosino2017}.
The observation was made with the Silicon Fast Astronomical Photometer and Polarimeter \citep[SiFAP2;][]{Meddi_2012PASP, Ambrosino2016, Ghedina2018}, currently mounted at the 3.58-metre INAF TNG (\citealt{Barbieri_1994SPIE}) at the Roque de Los Muchachos Observatory in La Palma. Using Silicon Photo-Multipliers (SiPMs), SiFAP2 is capable of counting individual optical photons, recording their detection times with a relative time resolution of 8 ns until 2024, now lowered to 80 ps. 

All SiPMs suffer from correlated noise, specifically crosstalk and afterpulsing \citep[][]{Klanner2019}, which induces a slightly higher normalization of the observed PDS with respect to what would be expected from Poissonian noise; this effect is of no practical consequence in the case of powerful sources, but can bring noise fluctuations closer to or slightly above detection thresholds, thus making detections of weak signals uncertain if not taken into account. 
In Appendix~\ref{app:rumore} we carry out a detailed study of this correlated noise and its impact on PDS, and we also provide two possible methods to properly include their effects during data analysis, ensuring that all the considerations in Sects.~\ref{sec:sign} and~\ref{sec:upper_limits} still hold. In our case, this translates to a simple renormalization of the power observed at each frequency in the coherent step by a factor of $\sim$$1/1.15$.

\subsection{Analysis setup}

The first detection of coherent pulsations from \die\ was obtained in 2009 in the radio band and led to measurements of its orbital parameters being refined to exquisite precision \citep{archibald2009}.
However, to properly validate our algorithm in the same conditions of its intended use, we decided to disregard all knowledge of the system obtained after the discovery of radio pulsations. Thanks to high-quality radial-velocity curves of the secondary star, \citet{Thorstensen2005} had previously already measured the orbital period to sub-second precision ($0.1980959 \pm 4 \cdot 10^{-7} $ days), and therefore we kept it fixed. They had also measured the radial velocity of the companion star as $K_2 = 268 \pm 4$~km/s$^{-1}$ (at a 1-$\sigma$ confidence level) from which the mass ratio, $q$, between the secondary and the primary star could be constrained to lie between $\sim$$0.04$ and $\sim$$0.75$ (at a 3-$\sigma$ confidence level). Through $a_\perp = K_2 \, q \, P_\mathrm{orb}/(2 \pi c)$, we obtained bounds on $a_{\perp}$ between 0.093 and 1.91~s. We also left $T_{\mathrm{asc}}$ unconstrained, letting it vary over the whole orbit. We then divided the SiFAP2 observation into 147 segments, each 128~s long, and ran the algorithm searching for pulsations at frequencies between 50 and 1550~Hz.

\begin{figure}[t!]
\centering
\resizebox{\hsize}{!}{\includegraphics{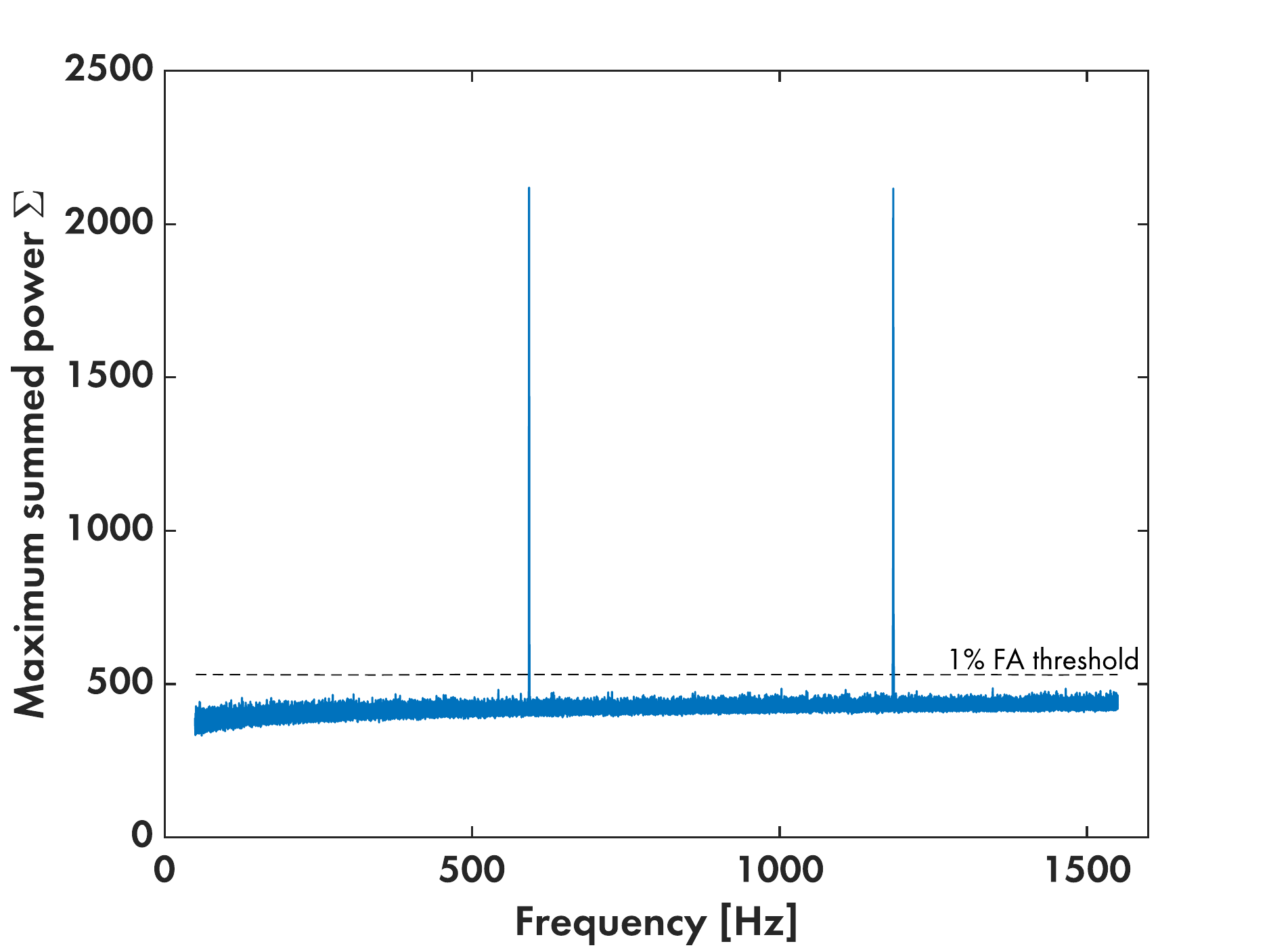}}
\caption{Maximum values of the summed power $\Sigma$, regardless of the corresponding orbital parameter combination, in the ${50 - 1550}$~Hz frequency range, on a SiFAP2 observation of \die. The dashed black line represents the $1\%$ false-alarm threshold, {which instead is calculated for all the combinations explored}.}\label{fig:1023}
\end{figure}

\subsection{Results}
Figure~\ref{fig:1023} shows the observed maximum value of the summed power, $\Sigma$, at each frequency searched. Two outstanding peaks ($\Sigma_1 = 2119.4, \Sigma_2 = 2116.2$) lie exactly at the frequencies of the first and second harmonics of the pulsar, the first peak being found in the bin corresponding to $592.4219(39)$ Hz. This value agrees with that obtained by \citet[][]{Burtovoi2020} using the known radio orbital ephemeris to correct optical data taken during those same days $f = {592.42146750 \pm 7\cdot10^{-8}}$~Hz.
The other two parameters in the combination corresponding to the first peak are $a_\perp = 0.3433 \pm 0.0068$~s and $T_{\mathrm{asc}} = {58878.991043 \pm 29\cdot10^{-6}}$~MJD, where the uncertainties we report are each equal to half their relative step in the parameter grid. They are compatible with the respective values known from literature, $a_\perp = 0.343356$~s \citep[][]{Jaodand2016} and $T_{\mathrm{asc}} = 58878.9911009 \pm 12 \cdot10^{-7}$ MJD \citep[][]{Illiano2023}. 

{We conclude that our semi-coherent algorithm would have been able to discover the pulsar using SiFAP2/TNG optical data alone, i.e. without making use of any previous pulse detection in the radio and in the X-ray band, and to determine its orbital parameters correctly.}

\section{Scorpius X-1} \label{sec:SCOX1}

\begin{table*}[t!]
\centering
\caption{Summary of \sco\ observations with SiFAP2/TNG}\label{tab:SCOobs}
\begin{tabular}{ccccrcc}
\hline
\hline
\textbf{Start date (UTC)} & \textbf{PI} & \textbf{Exposure [s]} & \textbf{MJDREF} & \multicolumn{1}{c}{$\mathbf{N_\gamma}$} & \textbf{Background rate [cts/s]} & \textbf{Passband filter} \\ \hline
2018/07/18    & Papitto & 3297.8         & 58317.94516     & 727\,115\,361                           & 28\,930         & White              \\
2018/07/19    & Papitto & 2397.8         & 58318.02016     & 324\,755\,950                           & 15\,236         & White              \\ \hline
2019/03/01    & Motta   & 3298.3         & 58543.13266     & 772\,856\,836                           & 11\,247         & White              \\
2019/03/01    & Motta   & 1798.2         & 58543.17433     & 435\,249\,516                           & 12\,506         & White              \\
2019/03/01    & Motta   & 3298.3         & 58543.19863     & 900\,960\,703                           & 13\,027         & White              \\
2019/03/01    & Motta   & 3298.3         & 58543.24516     & 1\,006\,292\,476                        & 16\,324         & White              \\ \hline
2019/04/10    & Papitto & 3124.2         & 58583.07884     & 429\,986\,263                           & 6\,496          & White              \\ \hline
2019/06/04    & Papitto & 3598.0         & 58638.08058     & 1\,502\,582\,911                        & 12\,181         & White              \\
2019/06/04    & Papitto & 3598.0         & 58638.12468     & 1\,323\,457\,013                        & 12\,073         & White              \\ \hline
2022/06/28    & Papitto & 8219.6         & 59758.90069     & 3\,974\,900\,556                        & 6\,354          & White              \\
2022/06/29    & Papitto & 7945.6         & 59759.00178     & 2\,759\,500\,899                        & 7\,060          & White              \\ \hline
2023/04/19    & Ambrosino & 6840.4         & 60053.13958     & 2\,060\,187\,287                        & 15\,511         & White              \\
2023/04/23    & Ambrosino & 7200.4         & 60058.13611     & 741\,060\,012                           & 6\,204          & Johnson-Bessel B              \\ \hline
\end{tabular}
\end{table*}

\begin{table*}[t!]
\caption{Setup and results of searches on \sco}              
\label{tab:SCOres}
\centering       
\begin{tabular}{c c c}   
\hline\hline       
\textbf{Dataset} & \textbf{D1 (2018 -- 2019)} &\textbf{D2 (2022 -- 2023)} \\   
\hline      
    $ f \, $[Hz]  & $50 - 1550$ & $50 - 1550$ \\
    $P_{\mathrm{orb}} \, $[s]  & $68023.92^\dagger$ & $68023.92^\dagger$\\
    $ a_\perp\, $[s]  & $1.45 - 3.25$ & $1.45 - 3.25$ \\
    $T_{\mathrm{asc}} \,$[MJD] &  $58638.1630 - 58638.1671$  & $ 60058.4761 - 60058.4824$   \\
    $T_{\mathrm{seg}} \, $[s]  & 512  & 512 \\      
    $M$               & 51   & 58  \\     
    $N_{\gamma,\mathrm{tot}}$  & 7\,058\,143\,860 & 9\,397\,338\,024\\
    $n^{\mathrm{trials}}$ & 4\,839\,772\,160 & 10\,552\,174\,592 \\
    $\Sigma^{1\%}$    & $234.27$   & $257.52$ \\
\hline
    $\Sigma_\mathrm{max}$        & $218.09$   & $236.14$ \\      
    $f(\Sigma_\mathrm{max}) \,$[Hz]     & $774.0371$ & $1296.1914$    \\
    $A_{\mathrm{ul}}$ & $9.76 \cdot 10^{-5}$ & $9.23 \cdot 10^{-5}$     \\
\hline                    
\end{tabular}
\tablefoot{  \centering $^\dagger$ The orbital period was kept fixed at its fiducial value. 
}
\end{table*}

SiFAP2 observed \sco\ during four campaigns that took place between 2018 and 2023, accumulating $\sim$16 hours divided into 13 observations, and $\sim$$1.7 \times 10^{10}$ photons (see Table \ref{tab:SCOobs}). All observations were taken in white light covering the 320-900 nm band \citep[see supplementary Fig. 1 in ][]{Ambrosino2017}, except for the last observation (April 23, 2023), which was conducted with a Johnson-Bessel B passband ($\lambda_{eff} = 445$~nm, $\Delta \lambda_{\mathrm{FWHM}} = 94$~nm). We corrected the photon detection times to the Solar System barycentre with the same pipeline used for the analysis in Sect.~\ref{sec:1023}, taking the source position from the Gaia catalogue: ${\alpha = 16^h19^m55.06927^s}$, ${\delta = -15^\circ38\arcmin25.01767\arcsec}$ \citep{GaiaCollaboration2020}. The background rates were quite different across observations, but always contributed less than 15\% of the total count rate.

\subsection{Analysis setup} \label{sec:SCOsetup}

The orbital parameters of \sco\ have long been monitored \citep[starting from][]{Gottlieb1975}, and there is an ongoing programme to periodically update them especially for CW searches \citep[][]{Galloway2014,Wang2018,Killestein2023}. 
The parameter that suffers from the largest uncertainties is $a_\perp$, whose value was poorly constrained owing to the lack of precise bounds on the system's masses: the most likely range is $(1.45 - 3.25) \,$s, as estimated in \citet[][]{Wang2018}. 
\citet[][]{Killestein2023} measured an orbital period of $P_{\mathrm{orb}}  = (68023.92 \pm 0.02) \, \mathrm{s}$ and a time of passage at the inferior conjunction of the companion star {$T_{\mathrm{con}} = 2456723.3272 \pm 0.0004$ BJD (UTC)} $ = {56722.8272 \pm 0.0004}$~MJD.\footnote{We note that the time in GPS seconds for $T_{\mathrm{con}}$ reported in \citet[][]{Killestein2023} differs by 4~s from the one in BJD (UTC). However, it was due to a typo, and the BJD value is the proper one (T. Killestein, private communication).} We then obtained a starting value of $T_{\mathrm{asc, old}} = T_{\mathrm{con}} - P_{\mathrm{orb}}/4 = (56722.8272 - 68023.92/86400/4)$ MJD $\pm \, 33$~s = $56722.6304$ MJD $\pm \, 33$~s for the epoch of passage of the compact object at the ascending node of the orbit. 
We extrapolated the value of $T_{\mathrm{asc}}$ closest to each of our observations as 
\begin{equation} \label{eq:tascnew}
T_{\mathrm{asc},i} = T_{\mathrm{asc, old}} + \left\lfloor \frac{\mathrm{MJDREF}_{i} - T_{\mathrm{asc, old}}}{P_{\mathrm{orb}}}\right\rceil \, P_{\mathrm{orb}} \ ,
\end{equation}
and their uncertainties through the usual addition in quadrature, giving
\begin{equation} \label{eq:tascunc}
\sigma_{T_{\mathrm{asc},i}} = 3\sqrt{(\sigma_{T_{\mathrm{asc, old}}})^2 + \left( \left\lfloor \frac{\mathrm{MJDREF}_{i} - T_{\mathrm{asc, old}}}{P_{\mathrm{orb}}}\right\rceil\, \sigma_{P_{\mathrm{orb}}}\right)^2} \, , 
\end{equation}
where $\lfloor \cdot \rceil $ indicates rounding to the nearest integer and $\mathrm{MJDREF}_{i}$ is the reference time of each observation; we used three times the 1-$\sigma$ uncertainty for both parameters. 
Considering these parameter ranges, we decided to take 512-s long segments to balance the load on our hardware and the amplitude we expected to reach with \sco's count rate (see Eq.~\ref{eq:Aul}). 

Among the strengths of semi-coherent searches is the possibility to jointly analyse observations quite far removed from one another, although this implicitly assumes a sufficient stability in the system's orbital configuration among observations. If present, a derivative on the orbital period, $\dot{P}_{orb}$, would translate into a change in $T_{\mathrm{asc}}$ as well. The value of $\dot{P}_{orb}$ can be estimated through \citep[see e.g.][]{diSalvo2008}
\begin{equation} \label{eq:dotP}
\frac{\dot P_{\mathrm{orb}}}{P_{\mathrm{orb}}} = 3 \left(\frac{\dot J}{J_{orb}} - \frac{\dot M_2}{M_2} \; g(\beta,q,\alpha)\right) \, ,
\end{equation}
where $g(\beta,q,\alpha) = 1 - \beta q - (1-\beta) (\alpha + q/3)/(1+q)$, $\beta$ is the fraction of the mass lost by the companion star that accretes onto the NS, $q = M_2/M_1$ is the mass ratio between the companion and the NS, and $\dot{J} / J_{orb}$ represents possible losses of angular momentum from the system. 

Supposing $\dot{J}$ to be caused by GW emission and assuming conservative mass transfer, the expected maximum $\dot{P}_{orb}$ can be estimated by taking the smaller of the two terms in Eq.~(\ref{eq:dotP}) to be zero. Here, $\dot M_2 = - RL_X/GM_1 = 3\times 10^{-8} \, M_\odot/\mathrm{yr}$, while $M_2 = (0.7_{-0.3}^{+0.8}) \, M_\odot$ \citep[at the $95\%$ c.l.;][]{Wang2018}.
We see that the dominant term is $-\dot M_2 / M_2$, and, assuming that $\dot J / J_{orb} = 0$,
\begin{equation}
\label{dotPlimit}
\frac{\dot P_{\rm orb}}{P_{\rm orb}} \le 3 \left(- \frac{\dot M_2}{M_2} \;
g(1,q,\alpha)\right) \, .
\end{equation}
The corresponding shift in $T_{\mathrm{asc}}$ is \citep[e.g.][]{Papitto2005}
\begin{equation} \label{eq:Deltatasc}
   \Delta T_{\mathrm{asc}}^{\dot P_{\mathrm{orb}}} = \left( \, \left\lfloor \frac{T_{\mathrm{span}}}{P_{\mathrm{orb}}}\right\rceil \, \right)^2 \, \frac{\dot P_{\mathrm{orb}} P_{\mathrm{orb}}}{2} = N_{orb}^2 \frac{\dot P_{\mathrm{orb}} P_{\mathrm{orb}}}{2}   \, ,
\end{equation}
which in our case gives $\Delta T_{\mathrm{asc}}^{\dot P_{\mathrm{orb}}} \sim 3.55\,(T_{\mathrm{span}}/\mathrm{yr})^2 \,$s. Since the smallest value of $\delta T_{\mathrm{asc},i}$ across our observations is $\sim$$11.4$ s, we decided to perform our searches by joining the observations of subsequent years into two datasets, 2018-2019 (D1) and 2022-2023 (D2), obtaining 51 and 58 segments, respectively. 

Given this value of $T_{\mathrm{seg}}$ and the other uncertainties of the parameters,  the step in $P_{\mathrm{orb}}$ would be greater than its 3-$\sigma$ c.l. uncertainties at all frequencies. Therefore, we only searched over $a_\perp$ and $T_{\mathrm{asc}}$. 
The first half of Table~\ref{tab:SCOres} summarizes the parameter ranges searched for both datasets, together with the conservative total number of trials, which is the sum of the number of parameter combinations at each frequency interval times the number of frequencies in each interval.

\subsection{Results} \label{sec:SCOresults}

We set our detection threshold as the power level that had a 1\% probability of being exceeded by counting noise alone. Given the number of segments of the semi-coherent steps and the number of trials considered (see Table~\ref{tab:SCOres} and Eq.~\ref{eq:sigmapfa}), this results in values of $\Sigma^{1\%}$ equal to 234.27 and 257.52, respectively. No significant coherent pulsation was found in the two datasets we analysed.\footnote{In D1, a spurious peak was found at 78 Hz, due to instrumental interference connected to the control loop of the telescope's tracking system. This was later rectified in 2019.}
The observed values of $\Sigma$ in the two datasets which give the respective upper limits on amplitude are given in the second half of Table~\ref{tab:SCOres}. The lowest is found in D2, with a value of $\Sigma_\mathrm{max} = 236.14$ at a frequency equal to $\sim$$1296.1914$~Hz. Through Eq.~(\ref{eq:noncentralcdf}) and (\ref{eq:Aul}), $\Sigma_\mathrm{max}$ can be translated into an upper limit on the background-subtracted pulse amplitude equal to $9.23 \cdot 10^{-5}$, at a 1\% false-alarm probability and 1\% false-dismissal probability. 

\section{Discussion and conclusions} \label{sec:discussion}

We carried out a semi-coherent search for optical pulsations over observations of \sco\ taken between 2018 and 2023 with the SiFAP2 instrument mounted at the TNG. No signal was found, resulting in the tightest upper limit on the pulsed fraction of the emission from the source, standing at $9.2 \cdot 10^{-5}$, at a 1\% false-alarm probability ($\mathcal{P}_\mathrm{FA}$) and 1\% false-dismissal probability ($\mathcal{P}_\mathrm{FD}$) or $8.4 \cdot 10^{-5}$ at a 1\% $\mathcal{P}_\mathrm{FA}$ and 10\% $\mathcal{P}_\mathrm{FD}$. This represents an improvement by a factor of four on the previous upper limit, which was set at $3.4 \cdot 10^{-4}$ (at 1\% $\mathcal{P}_\mathrm{FA}$ and 10\% $\mathcal{P}_\mathrm{FD}$), through a similar search on Rossi X-ray Timing Explorer data in the 3--60~keV X-ray band by \citet[][]{Galaudage2022}. The approximately one-order-of-magnitude increase in the count rate observed in the optical band compared to in X-rays is the main driver of such an improvement.

The lack of coherent pulsations in optical observations of \sco\ that we present in this paper aligns with the non-detection of pulsations from most accreting NSs in LMXBs \citep[148 out of the 188 NS systems in the \mbox{XRBcats} catalogue;][]{Avakyan2023,Niang2024}, with upper limits on the X-ray pulsed amplitude generally lower than 1\% \citep{vaughan94,dib2005,Messenger2015,Patruno2018,Galaudage2022}. Most of the observed properties suggest that the lack of pulsations can be ascribed to the absence of a magnetosphere (see e.g. the discussion in \citealt{Patruno2018} and \citealt{patruno2021}). 
Magnetic screening by the infalling matter on the NS quenches the surface magnetic field \citep{Bisnovatyi-Kogan1974,Cumming2001} and would naturally explain why only systems accreting at lower rates (such as AMSPs) show X-ray pulsations. Scattering of pulsed X-rays by optically thick clouds is an additional mechanism that could explain the lack of pulsations in the brightest accreting systems \citep{Brainerd1987,Kylafis1987,Titarchuk2002}. 
These scenarios, the first one especially, could completely prevent pulsations from arising.
An arguably more optimistic alternative is that \sco\ will eventually show pulsations through the same, still unclear, mechanism that brought a different LMXB system, Aql X-1, to display a strong ($6\%$ in pulsed fraction), coherently pulsated signal for a very brief interval ($\sim$150 s) over hundreds of hours of on-source time \citep{Casella2008}.

To reach such a low upper limit on the pulsed fraction, we applied semi-coherent strategies to optical data for the first time. 
We decided to employ these strategies owing to their cost-effectiveness in the Fourier analysis of long observations of systems for which limited knowledge on the orbital parameters is available, building on the method described by \citet{Messenger2015}. 
These techniques limit the coherent analysis to small segments of the whole observation, generally through separate PDS, to subsequently sum the relevant statistics across segments while keeping track of the expected orbital evolution of the signal \citep{Messenger2011,Messenger2015}.
Gamma-ray astronomers were the first to adopt semi-coherent searches in the analysis of electromagnetic data, owing to the need to integrate over long periods to obtain a significant number of gamma-ray photons \citep{Pletsch2012Science}. 
Through the massive volunteer computing \emph{Einstein@Home} project \citep{Knispel2010}, these searches successfully found three previously unknown gamma-ray pulsars in binaries \citep{Pletsch2012Science,Nieder2020detection,Clark2021}. 
The main difference between the method used in gamma-ray searches and the one employed here is that, rather than using Fourier PDSs, the former is based on the time-differencing technique developed by \citet{Atwood2006}, which calculates autocorrelations between photon detection times no more than a given interval apart \citep[see e.g.][]{Pletsch2014}. This is very efficient in the gamma-ray band precisely because of the low number of photons involved, and had already been successful in the search for isolated gamma-ray pulsars \citep[see e.g.][]{Abdo2009,SazParkinson2010,Clark2017}. However, the high number of photons at other wavelengths makes time-differencing unfeasible: to give an example, the timing analysis in \citet{Clark2021} included 6571 gamma-ray photons over 11 years of observations, while the photons in the 19-ks optical observation of \die\ analysed in Sect.~\ref{sec:1023} amounted to $\sim$$4.3\times10^8$.
Therefore, at other bandwidths it is more efficient to divide the observation into segments and produce a PDS for each possible set of observed frequency and its derivatives allowed by the orbital uncertainties: that is the first, coherent step of the algorithm employed in this paper.
Doing so is essentially equivalent to performing acceleration or jerk searches, depending on the number of derivatives used, $s^*$. Since this step is to be carried out anyway, this implies that the results of acceleration/jerk searches are produced at no additional cost while running this algorithm. Although they are less efficient than the full, semi-coherent search in the case of weak but persistent signals, they can instead detect slightly stronger but transient pulsations that are present only in one or a few segments and could otherwise be missed in the sum over a long observation. Therefore, checking the results of the coherent step separately can help identify transient pulsations such as the ones commonly found in spider systems in between their long eclipses \citep[e.g.][]{Thongmeearkom2024} and the aforementioned one in Aql X-1 \citep{Casella2008}.

The semi-coherent step only involves nearest neighbour searches over the coherent space for each of the physical parameter combinations in the semi-coherent one in each segment; the powers in the frequency band of interest are then summed across all segments. To create a grid of combinations of physical parameters, we employed the exact expressions described by \citet{Caliandro2012} to define a minimum distance in each dimension of the semi-coherent space (similar, analytical formulations were given in e.g. \citealt{Dhurandhar2001}, \citealt{Messenger2011}, \citealt{Leaci2015}, and \citealt{Nieder2020method}).
The way we tiled both the coherent and semi-coherent spaces of the algorithm ensures that the fraction of signal power recovered is always at least $0.95 \cdot 0.975^r$ overall, where $r$ is the number of orbital parameters to be searched. 

If the orbital period is known to sufficient precision such that it can be kept fixed, a further improvement can be implemented: since the orbital modulation of the signal is periodic, one could in principle choose the segment length $T_\mathrm{seg}$ to be a divisor of $P_{\mathrm{orb}}$ so that the nearest-neighbour search only needs to be carried out $P_{\mathrm{orb}}/T_\mathrm{seg}$ times instead of $M$ times.\footnote{As always, care should be exercised to ensure that the Nyquist frequency used for the PDS is tuned to the resulting $T_\mathrm{seg}$ in order to have well-factorized arrays, otherwise the slowdown of the FFTs in the coherent step will vastly exceed this benefit (see e.g. \citealt{Frigo2005}, \citealt{Johnson2008}, Ch.~11, and \url{https://www.fftw.org/fftw2_doc/fftw_3.html} for FFTW.).}
In our searches, the semi-coherent step was sub-dominant with respect to the first one, so we decided to forgo this to better balance the load on our hardware in the coherent step.

We ran our algorithm on two custom computers equipped with 48-GB NVIDIA GPUs. Our code, written in MATLAB, exploits the GPU for the coherent step, and a mix of GPU and CPU for the semi-coherent one. For reference, after some testing and fine-tuning, the analysis of the \sco\ data described in Sect.~\ref{sec:SCOX1} took three weeks using a setup with a 48-GB NVIDIA video card, 96 AMD cores, and 1 TB of DDR4 RAM. However, we were mostly limited by memory constraints and will alleviate these through a more modular version of the code, to make it more flexible to operate on different setups.
As a comparison between methods, we note that running a fully coherent blind search just on our longest observation, the one from 2022/06/28, would have required a number of parameter combinations $\sim$240 times larger than the semi-coherent search over the whole D2 dataset, which would have been unfeasible on our machine, to reach a sensitivity of $\sim$$1.17 \cdot 10^{-4}$ in pulsed amplitude (at 1\% $\mathcal{P}_\mathrm{FA}$ and $\mathcal{P}_\mathrm{FD}$).

Through the test of the code on a SiFAP2 observation of \die\ we also proved that, even without the precise measurements of the systems' spin and orbital parameters made after the discovery of its pulsar in the radio band, our optical data alone would have been sufficient to detect the neutron star pulsation to a high level of significance. In fact, even during the coherent step, some segments were found to firmly show pulsations ({$\mathcal{P}_\mathrm{FA}< 10^{-5}$} after accounting for all the trials in the coherent step), which indicates that acceleration searches too would have been able to identify the pulsar without the information coming from radio observations. These results further cement the viability of the optical band in the search for pulsars.

Overall, this work represents an additional confirmation of the advantage of using semi-coherent techniques when searching for pulsations in binary systems, at all wavelengths, when limited by computational power. In the future, we plan to expand on our present results by including the effects of a possible derivative of the orbital period, which is currently too costly, in order to jointly analyse observations spanning many years. We will also explore different methods, such as mixed analyses in the frequency and time domain \citep[e.g.][]{Leaci2017}, to find the most efficient ones to go through our archival observations of faint candidate pulsar systems.

\begin{acknowledgements}
RLP wishes to thank Gian Luca Israel and Piergiorgio Casella for the useful comments and discussion. We would like to thank the anonymous referee for the useful comments which improved this work. We acknowledge support from the Italian Ministry of University and Research (MUR), PRIN 2020 (2020BRP57Z, PI: Astone) \textit{Gravitational and Electromagnetic-wave Sources in the Universe with current and next-generation detectors (GEMS)}, INAF Research Large Grant \textit{FANS} (PI: Papitto), INAF Guest Observer Grant \textit{PULSE-X} (PI: Papitto), Fondazione Cariplo/Cassa Depositi e Prestiti (Grant 2023-2560, PI: Papitto), and INAF Research Data Analysis Grant (Unveiling the secrets inside SiFAP2 data, PI: Ambrosino).
\end{acknowledgements}

\bibliographystyle{aa_url.bst}
\bibliography{bibo}

\appendix
\section{PDS corrections for correlated noise}\label{app:rumore}

Analysing any light curve from SiFAP2 in the 2017-2023 period, an overdispersion in the data with respect to a Poisson distribution is apparent. In most cases, the Leahy-normalized power density spectrum (PDS) shows an average power at high frequencies $\sim$$2.3$, which is significantly higher than the expected value of 2, were the data well approximated by Poissonian noise \citep{leahy83}. 
Such an overdispersion is common in systems relying on silicon photomultipliers (SiPMs) such as SiFAP2, and it is mainly due to the additional correlated noise introduced by afterpulsing and optical crosstalk (see e.g. \citealt{Acerbi2019}, \citealt{Klanner2019} and references therein). The former effect arises when, during an avalanche initiated by a photon hitting one of the detector pixels, charge carriers become trapped within the crystal lattice defects: their delayed release thus triggers a secondary avalanche on the same pixel. The latter instead is effected by photons emitted in an avalanche that trigger one or more avalanches in neighbouring pixels. As is the case for afterpulsing, the additional avalanches give rise to spurious events which can be indistinguishable from an actual signal.

The characterization of SiPMs has been extensively studied \citep[see e.g.][]{Klanner2019}, and the observed signal distribution arising from correlated noise has been well modelled by a generalized Poisson (GP) distribution \citep[][]{Vinogradov2009,Gallego2013,Chmill2017}: this represents the fact that while incoming photons follow Poissonian statistics, the number of pixels which are fired by each photon is not constant.\footnote{To avoid a long-standing ambiguity between \textit{compound}, \textit{mixed}, and \textit{generalized} probability distributions, in this paper we follow the naming convention used in \citet{Univariate_discrete_distributions},~Chap.~9.} 
In particular, \citet{Gallego2013} explored the fired-pixels distribution arising from various crosstalk types, taking into account possible saturation effects: each pixel of the array has a finite number of available neighbouring pixels to excite via crosstalk, and as crosstalk propagates the number of unfired pixels in the neighbourhood that can be affected slowly decreases. The authors give a complete description of the probability distribution of pixels fired by a photon hitting a pixel, for the cases in which each pixel can affect only one pixel in its vicinity, its four nearest neighbours, or its 8 nearest neighbours.

For a given intrinsic distribution of primary events (i.e. pixels being fired by photons directly), $P_\mathrm{pr}$, the distribution of the total number of fired pixels we observe is:
\begin{subequations}\label{eq:totctdist}
\begin{align}
&P_{\rm tot}(0)=P_{\rm pr}(0) \qquad \\
&P_{\rm tot}(k)=\sum_{m=1}^{k}P_{\rm pr}(m)\,P_m(k)\qquad &\text{for } k\geq1\,.
\end{align}
\end{subequations}
Here $P_m(k)$ is the probability of $k$ pixels being triggered starting from $m$ primaries, which represents the additional noise given by crosstalk, and only depends on the number of neighbouring pixels $n$ that each fired pixel can excite and the total crosstalk probability $\epsilon_{ct}$ that any number of pixels will be excited by crosstalk effects. Exact and approximate expressions for $P_m(k)$ are given in Sect.~2 of \citet{Gallego2013}, to which we refer the interested reader.

Supposing the primary photon distribution to be Poissonian in nature, its statistics is entirely described by the mean $\mu$: we can then estimate both $\mu$ and $\epsilon_{ct}$ from the first two values of the observed count distribution through \citep[e.g.][]{Vacheret2011,Gallego2013, Univariate_discrete_distributions}
\begin{subequations} \label{eq:mupoiss}
\begin{align}
   & \mu = - \ln{P_{\rm tot}^\mathrm{obs}(0)} \\
   & \epsilon_{ct} = 1 - \frac{P_{\rm tot}^\mathrm{obs}(1)}{\mu \exp{\left[-\mu\right]}} \,,
\end{align}
\end{subequations}
and from those derive the $P_m(k)$ expected for that $\epsilon_{ct}$. We can then use Eq.~(\ref{eq:totctdist}) to construct the resulting expected total, crosstalk-affected distribution and compare it with the observed one at all $k$.
By analysing a few tens of SiFAP2 observations from 2017 to 2023, we extracted values of $\epsilon_{ct}$ which are on average $\sim$7$\%$, with small variations of $\sim$1$\%$ at times where the instrumental setup was changed. 

\begin{figure}[t!]
\includegraphics[width=0.5\textwidth]{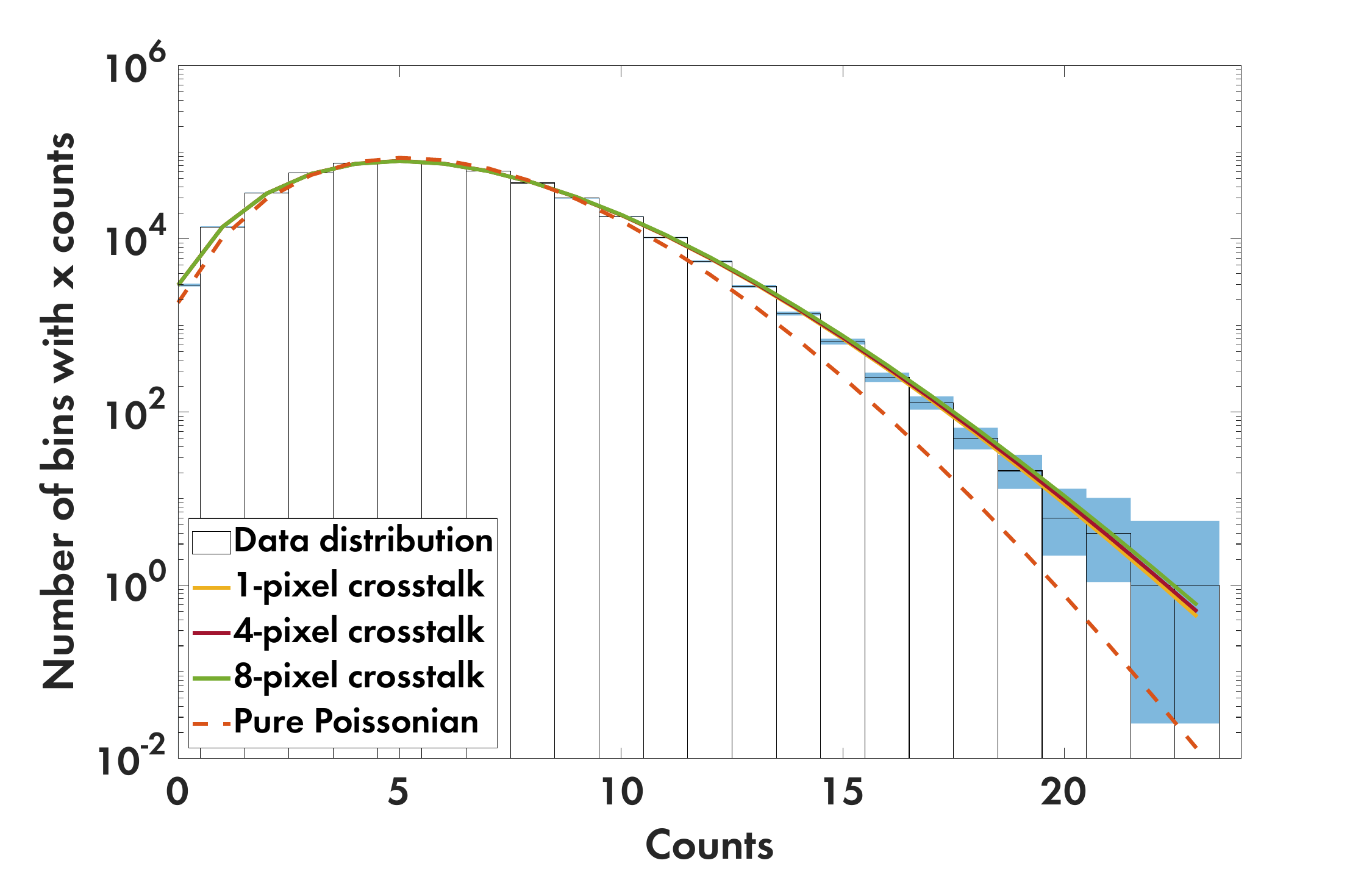}
\caption{Counts distribution from a 128-s SiFAP2 lightcurve binned at $2.5\times10^{-4}$~s (histogram), with binomial uncertainties (light blue area, 95\% c.l.), compared with the expected distributions from pure Poissonian noise (orange dashed line), and GP noise arising from 1-, 4-, and 8-pixel crosstalk (solid lines).}\label{fig:noisecountdist}
\end{figure}

For all of them, we checked that the crosstalk-affected GP distribution we expected agreed well with the data. Figure~\ref{fig:noisecountdist} shows one such example, for a 128-s-long segment of the observation of \die\ made on 2020/01/30: the histogram represents the distribution of pixel counts within $0.00025$-s bins, with binomial 95\% confidence levels for the underlying unknown distribution indicated by the light-blue shaded area \citep[][]{Clopper1934}. The orange dashed line represents the distribution expected from pure Poissonian noise with average equal to the observed count average, while the solid lines indicate the expected distributions in the case of crosstalk-modified Poissonian noise, assuming 1-, 4-, and 8-pixel crosstalk: the agreement is very good for all three models, which, for our purposes, are virtually identical.

Moreover, we can also derive the expectation value and variance of the total distribution \citep[e.g.][]{Univariate_discrete_distributions,Gallego2013}:
\begin{subequations}\label{eq:GPprob}
\begin{align}
&E_{\rm tot}=E_{\rm pr}\,E_1 = \mu\,E_1\label{eq:etot}\\
&\sigma^2_{\rm tot}=E_{\rm pr}\,\sigma^2_1+\sigma^2_{\rm pr}\,E^2_1 = \mu\left( \sigma_1^2 + E_1^2\right)\,,\label{eq:vartot}\\
&\qoppa = \frac{\sigma^2_{\rm tot}}{E_{\rm tot}} = \frac{\sigma_1^2 + E_1^2}{E_1^2}\, ,\label{eq:qoppa}
\end{align}
\end{subequations}
where $\sigma_1^2$ and $E_1$ refer to the distribution of pixels excited by a single photon and only depend on $\epsilon_{ct}$ \citep[see][]{Gallego2013}; $\qoppa$ is the total dispersion index (in this case, equal to the overdispersion with respect to a Poissonian since for it $E = \sigma^2 = \mu$).

\begin{figure}[t!]
\includegraphics[width=0.5\textwidth]{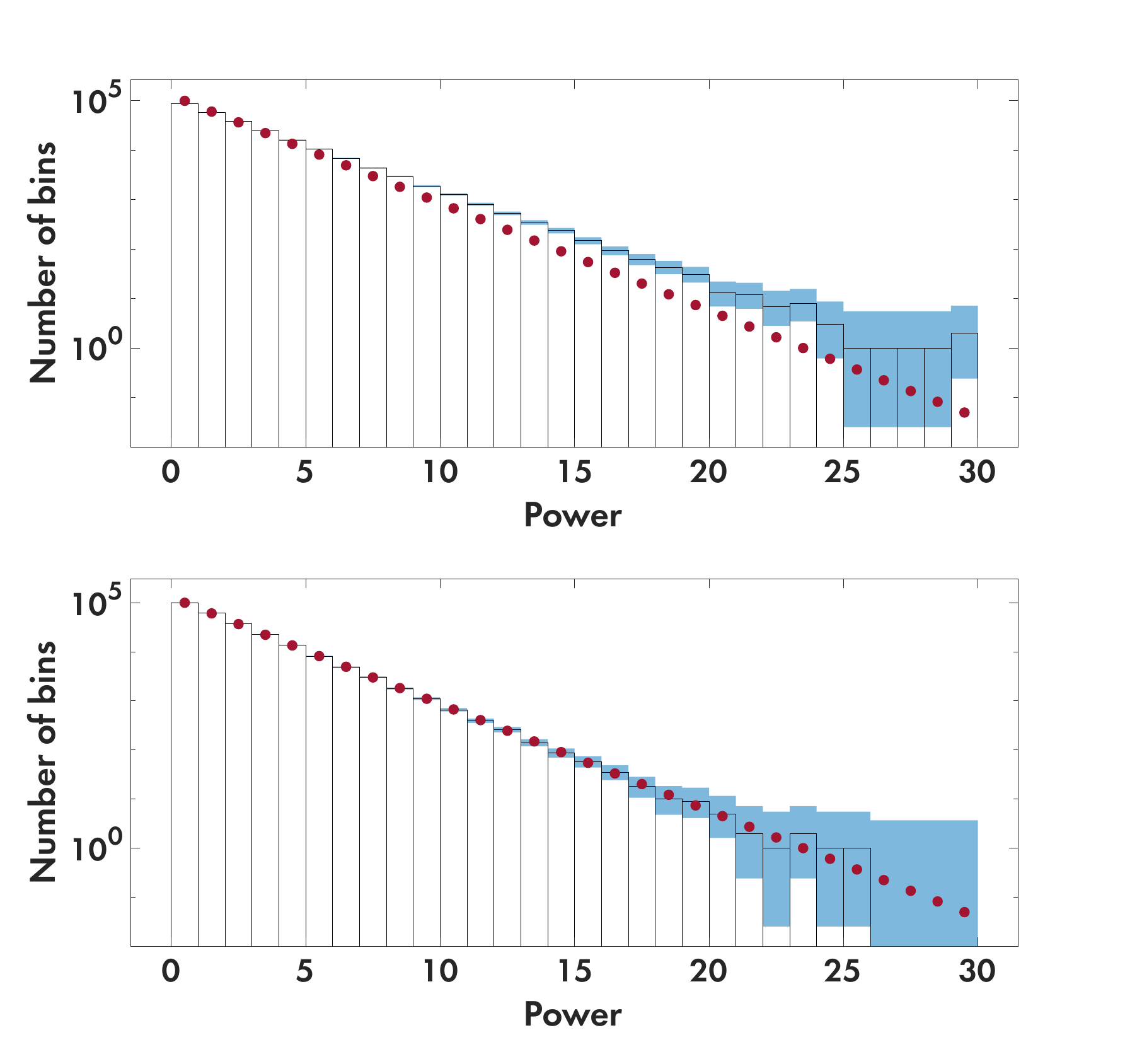}
\caption{\textit{Upper panel -} Power distribution of the PDS from the same data in Fig.~\ref{fig:noisecountdist} (histogram), with binomial uncertainties (light blue area, 95\% c.l.), compared with a $\chi^2$ distribution with two d.o.f. (burgundy points, normalized to the total number of powers). \textit{Lower panel -} Same as upper panel, but after rescaling every power in the PDS by the dispersion index of the data. The $\chi^2$ distribution in the two panels is the same.}\label{fig:noisepowdist}
\end{figure}

Since our noise is well described by a GP distribution, we can adjust our analysis in one of two ways. 
The first method would be considering that as the expected photon distribution and revisiting all statistics and thresholds according to Eq.~(\ref{eq:GPprob}): the advantage of this method is that, when sure of the cause of the correlated noise, it can be used even outside PDS-based analysis (as in, e.g., Campa et al., in preparation). 
The second one is to exploit some of the properties of the noise distribution to still follow the usual approach, after a renormalization: if the noise tends to a white noise, the PDS can simply be renormalized by its dispersion index \citep[see e.g.][]{israel96}. This ``tendency to white noise'' of a distribution is formally proven if the sample mean of variables following that distribution tends to a Gaussian one. For generalized shot noise processes, including GP ones, a central limit theorem towards Gaussianity was demonstrated by \citet{Rice1977}. However, even if the noise were described by a different distribution, the range of distributions which tend to Gaussianity is quite extensive, including all finite-variance ones \citep[the classical central limit theorem; see e.g.][]{FellerVol2}, all log-concave distributions \citep[][]{Klartag2007}, or equivalently strongly unimodal ones \citep{Ibragimov1956,Keilson1971}, and more in general all the ones whose truncated second moment function varies slowly \citep[e.g.][Chap.~IX]{FellerVol2}. The wider range of applicability of this second method makes it more robust to possible small differences of the real instrumental noise with respect to a given model. 

{Therefore, the following approach can be of use for PDS analysis whenever correlated noise is found in the data:
\begin{itemize}
    \item Check that the noise follows a distribution which tends to Gaussianity, in the sense described above;
    \item If so, renormalize the PDS by dividing all powers by the dispersion index $\qoppa$ of the data, or, equivalently, half of the average high-frequency power;\footnote{Were the noise not white, one can instead use any of the prescriptions for coloured noise, such as the one in \citet{israel96}.}
    \item Carry out the PDS analysis on the renormalized spectra following the usual Leahy-normalized thresholds and probability assumptions, as in Sect.~\ref{sec:sign} and~\ref{sec:upper_limits}.
\end{itemize}}
As an example of this, the upper panel of Fig.~\ref{fig:noisepowdist} shows the distribution of powers (histogram) of the Leahy-normalized PDS of the same data from Fig.~\ref{fig:noisecountdist}, compared with a $\chi^2$ distribution with two degrees of freedom (normalized, burgundy points), which would be expected in the case of Poissonian noise \citep[][]{leahy83}: the instrumental power distribution shows a significantly shallower slope than expected. If instead we first divide all power values by $\qoppa$, we obtain the distribution in the lower panel: the agreement with the $\chi^2$ distribution, which is identical in the two panels, is now restored.
In the present paper, we used this approach for all of the analysis described in the main text.

\section{Algorithm summary}\label{app:algo}

In order to summarize the algorithm presented in Sect.~\ref{sec:method} and help in the reproducibility of our results, we present here an outline of the whole semi-coherent search discussed in the main text.

\begin{algorithmic} 
  \HEADER{Semi-coherent search}
  \FOR{each frequency band considered}
    \HEADER{Coherent step}
      \STATE From the minimum of $P_{\mathrm{orb}}$ and the maximum of $a_\perp$ obtain the limits on each dimension in the coherent space through Eq.~(\ref{eq:nis})
      \STATE Find the largest order $s^*$ to contribute, for the chosen maximum mismatch $\mu^*$, to the Taylor expansion through Eq.~(\ref{eq:deltanis})
      \STATE Create a coherent template bank with all possible ${\vec\nu = (\nu_1,\nu_2,\dots,\nu_{s^*})}$ through Eq.~(\ref{eq:deltanis})
      \FOR{each segment $m$}
        \FOR{each $\vec{\nu}$}
            \STATE Compute $\vec{\tau}^{(m)}(\vec{\nu})$ from the photons' detection times through Eq.~(\ref{eq:tautaylor})
            \STATE Compute the PDS on the $\vec{\tau}^{(m)}(\vec{\nu})$, obtaining $\Lambda^{(m)}$
        \ENDFOR
      \ENDFOR
    \ENDHEADER
    \HEADER{Semi-coherent step}
      \STATE Create a semi-coherent template bank of combinations $\vec{\theta} = (f, P_{\mathrm{orb}}, a_\perp, T_{\mathrm{asc}})$
      \FOR{each segment $m$}
        \FOR{each $\vec{\theta}$}
            \STATE Compute {$\vec{\nu}(\vec{\theta})$} through Eq.~(\ref{eq:nis})
            \STATE Find nearest neighbour in the coherent space
            \STATE Sum the corresponding $\Lambda^{(m)}$ to $\Sigma(\vec{\theta})$
        \ENDFOR
      \ENDFOR
    \ENDHEADER
    \IF{$\max[\Sigma(\vec{\theta})] > \Sigma^{\mathcal{P}_\mathrm{FA}}$}
        \STATE There is a detection (at the false alarm c.l. $\mathcal{P}_\mathrm{FA}$; Eq.~\ref{eq:sigmapfa})
    \ELSE
        \STATE Find the $\Sigma(\vec{\theta})$ that maximizes the value of $A$ through Eq.~(\ref{eq:Aul}) to obtain the upper limit on the pulsed amplitude
    \ENDIF
  \ENDFOR 
  \ENDHEADER
\end{algorithmic}

\section{Significance thresholds for multiple dependent trials in \texorpdfstring{$\chi^2$}{chi-squared} random noise}
\label{app:multitrial}

Multi-trial searches inevitably entail facing the `multiple comparison problem', i.e. the fact that given enough different trials computed, one can surely find significant structures in any statistic if the number of trials itself is not taken into account \citep[see e.g.][]{Tukey1953,Hochberg1987}. Generally, procedures to do so put constraints on the familywise error rate (FWE), i.e. the chance of any false positives across all trials. In our case, we compare the summed power $\Sigma$ obtained correcting the data for the binary motion at each template $\vec{\theta} = (f, P_{\mathrm{orb}}, a_\perp, T_{\mathrm{asc}})$ with the distribution of the power expected in the case of noise alone, a $\chi^2_{2M}$ distribution. Our null hypothesis (i.e. that only counting noise is present over the whole search region) is therefore rejected at a $1-\mathcal{P}_\mathrm{FA}$ confidence level if any $\Sigma$ is above the threshold given by Eq.~(\ref{eq:sigmapfa}), which corrects for the number of trials $n$. The correction we used is the Dunn-\v{S}id\'{a}k one \citep{Sidak1967}, which sets the level of false alarm probability on the single test as:
\begin{equation}\label{eq:pdunnsidak}
    \mathcal{P}_\mathrm{FA}^{st} = 1 - (1-\mathcal{P}_\mathrm{FA})^{1/n} \, .
\end{equation}
This is commonly applied and for $\mathcal{P}_\mathrm{FA}$ values lower than $\sim$$5\%$ it is essentially equivalent to the Bonferroni correction, $\mathcal{P}_\mathrm{FA}^{st} = \mathcal{P}_\mathrm{FA}/n$, which represents an upper limit on the FWE \citep[see e.g.][Appendix 2]{Hochberg1987}.

\begin{figure}[t!]
\includegraphics[width=0.5\textwidth]{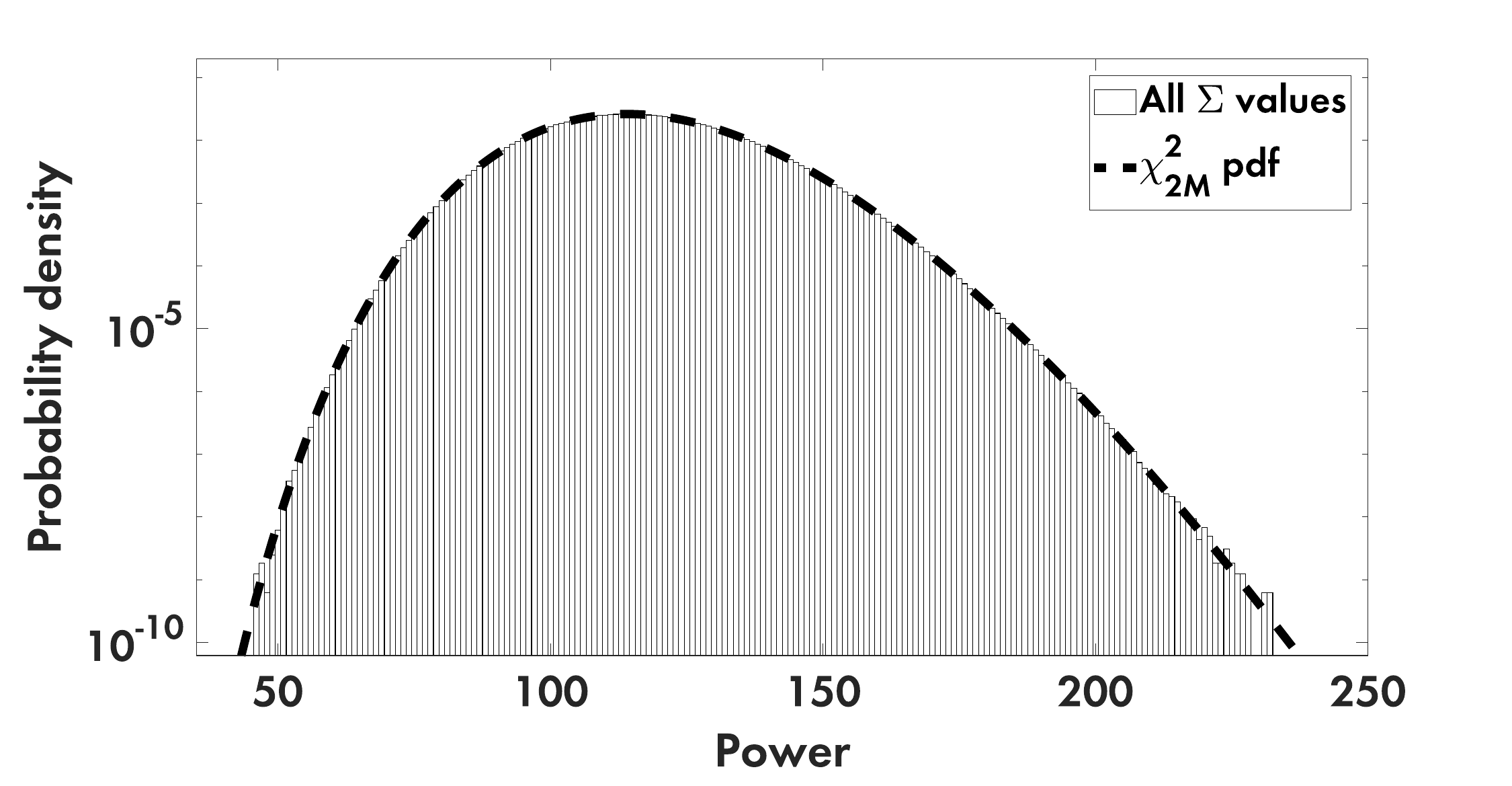}
\includegraphics[width=0.5\textwidth]{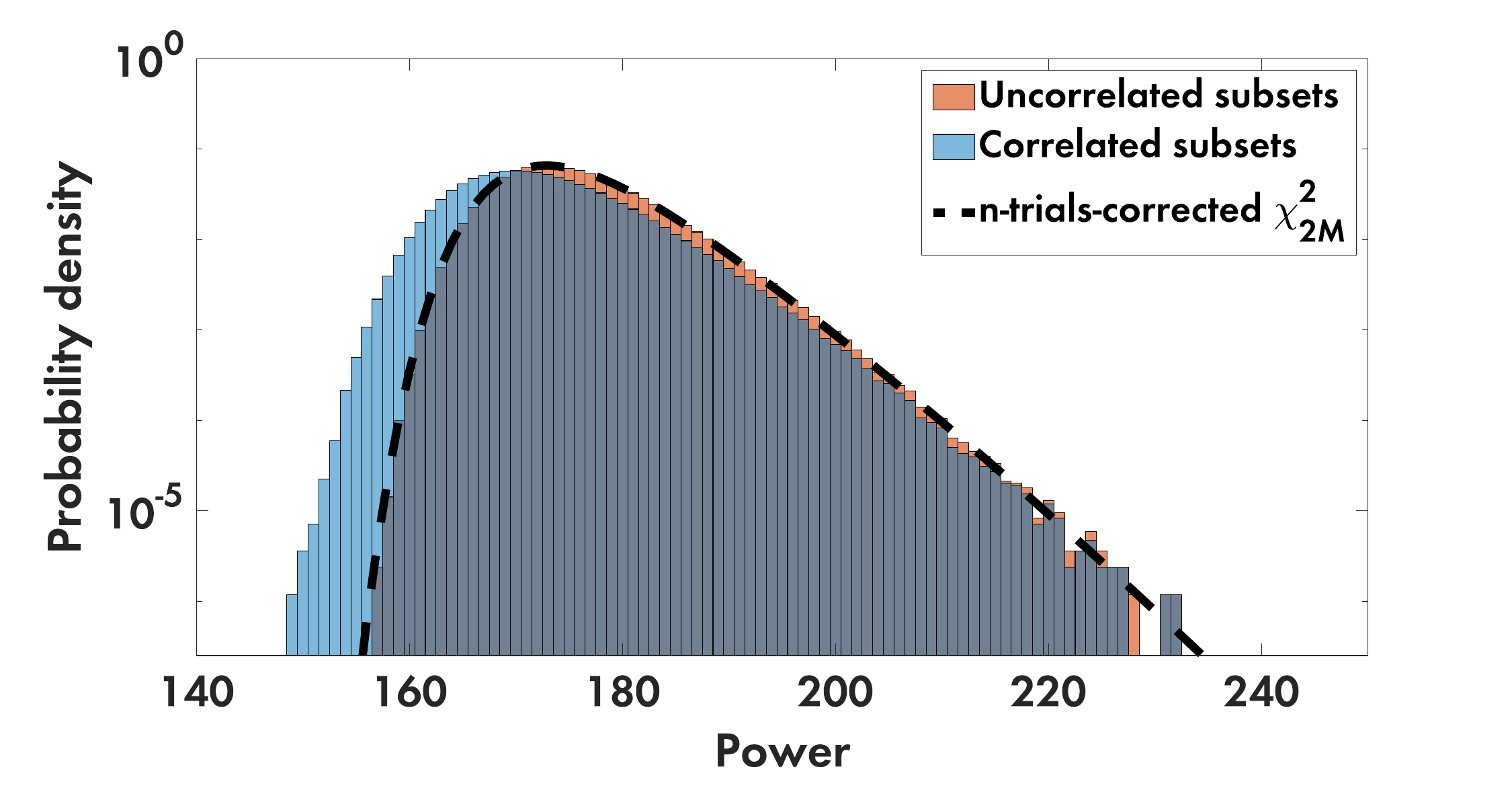}
\caption{\textit{Upper panel -} Power distribution of all of the $\Sigma$ values obtained in the analysis of the $1349-1450$~Hz interval for dataset D2 (see Sect.~\ref{sec:SCOX1}). The histogram was renormalized to the number of $\Sigma$ values considered, i.e., the $51712 \times695\times 45$ templates (over the $f$, $a_\perp$, and $T_{\mathrm{asc}}$ dimensions). The dashed line is the expected $\chi^2_{2M}$ probability density function in the case of noise alone.
\textit{Lower panel -} Distributions of the maximum values of $\Sigma$ over subsets of our parameter space. The blue histogram was obtained by taking non-overlapping blocks of close templates, with block sizes equal to 128, 3, and 5 points in the $f$, $a_\perp$, and $T_{\mathrm{asc}}$ dimensions, respectively; the orange histogram was obtained in the same way, but after randomly shuffling the position of all points in our space. Both histograms were renormalized to the number of subsets, $842\,340$. The dashed line is the expected $n$-trials-corrected $\chi^2_{2M}$ probability density function (Eq.~\ref{eq:multipdf}).
}\label{fig:distro_max}
\end{figure}

The Bonferroni bound is very general, and makes no assumption about the independence of the trials computed. If however the values of the statistic considered are correlated, it tends to be conservative \citep[e.g.][]{Nichols2003,Worsley2004}. Let us take as an example the summed powers $\Sigma$ obtained in Sect.~\ref{sec:SCOX1}: the upper panel in Fig.~\ref{fig:distro_max} shows the distribution of the $\Sigma$ values for dataset D2 over the whole binary parameter space for the frequency interval $1349-1450$~Hz, which is consistent with noise following a $\chi^2_{2M}$ distribution (here, $M = 58$). 
Nonetheless, since our template grid is dense to avoid potentially missing signals, the noise realizations at templates close to one another will not be completely independent: the photons' detection time corrections for close combinations of binary parameters will be similar, and that in turn will affect the PDS. We follow the approach used in the Appendix of \citet{Nieder2019} to study the distribution of the maxima within subsets of our data: since in the analysis we kept the orbital period fixed, we have a 3D parameter space in $f$, $a_\perp$, and $T_{\mathrm{asc}}$ which we divide in three dimensions in $404 \times 139 \times 15$ pieces, respectively. We thus obtain $842\,340$ non-overlapping subsets of our parameter space, and for each of those we find the maximum value value of $\Sigma$.
The distribution of these maxima is shown as the light blue histogram in Fig.~\ref{fig:distro_max}, lower panel, compared to the expected probability density of the $n$-trials-corrected $\chi^2_{2M}$ distribution (dashed black line) which can be obtained by taking the derivative of Eq.~(\ref{eq:pmulti}) with respect to $\Sigma$:
\begin{equation}\label{eq:multipdf}
    \dfrac{\mathrm{d\,} (1 -\mathrm{\mathcal{P}_n(\chi_{2M}^2\geq{\Sigma}))}}{\mathrm{d\,} \Sigma} = n\, \frac{2^{-M} \, e^{\Sigma/2} \, \Sigma^{M-1}}{\Gamma(M)} \, \left(P\left(M,\frac{\Sigma}{2}\right)\right)^{n-1} \, ;
\end{equation}
the number of trials in this case is $n = 1920$, i.e. the number of templates in each subset. As an additional comparison, the orange histogram represents the distribution of the $\Sigma$ maxima calculated over the same number of non-overlapping subsets of our parameter space, this time however building the subsets from randomly shuffled points: clearly doing so removes the correlations in the single subsets, and the distribution follows again what is expected from purely independent trials. 
Therefore, in the correlated case most of the distribution of the maxima of our $\chi^2$ noise is skewed towards lower powers, but gradually approaches the behaviour of the uncorrelated case for the highest powers. This is perfectly consistent with what is expected from random field theory, in which the $\chi^2$ field for non-independent noise realizations was first studied by \citet{Worsley1994}. Later studies yielded ways to quantitatively assess the statistical significance of fluctuations for that field \citep{Worsley1996,Worsley2001}: nevertheless, for high values of the $\chi^2$ statistic, the ones we are more interested in, the Bonferroni correction provides the tightest bound to the FWE and therefore to $\mathcal{P}_\mathrm{FA}^{st}$ \citep{Nichols2003,Worsley2004,Taylor2007}. This ensures that the sensitivity thresholds quoted across the rest of the paper are correct, at most being conservative. 
The same check was done at all frequency intervals for both \sco\ datasets, yielding the same results.

\end{document}